\documentclass{article}
\usepackage[utf8]{inputenc} 
\usepackage[T1]{fontenc}
\usepackage[english]{babel}
\usepackage{lmodern}
\usepackage{amsmath}
\usepackage[letterpaper,top=2cm,bottom=2cm,left=3cm,right=3cm,marginparwidth=1.75cm]{geometry}
\usepackage[autostyle]{csquotes}
\usepackage[backend=bibtex,style=numeric]{biblatex} 
\usepackage{graphicx}
\numberwithin{equation}{section}
\title{Capillary phenomena: New fundamental formula}
\author{Noureddine Djama\\University of Jijel, Algeria.\\
Department of Process Engineering.\\ E-mail: noureddine.djama@univ-jijel.dz}
\date{}
\begin{document}
\maketitle
\begin{abstract}
This study proposes a new fundamental formula that describes in a more coherent way, the rise and fall of liquids in capillaries. The variation of the contact angle classically associated with
these phenomena appears to be the indirect result of a more authentic physical parameter, which we call  the “apparent capillary range”. This range depends on factors expected to affect the contact angle, such as liquid-solid adhesion forces, liquid-liquid cohesion forces, liquid density, gravitational forces and the geometric shape of the capillary section. Our main objective in this work is not to criticize the classical theory—a task that has been largely accomplished—but to
present a more general and coherent approach, which perfectly reconciles the thermodynamic and mechanical points of view and makes the interpretation of various configurations clearer. This new perspective can serve as a platform to guide researchers’ efforts toward more promising results. In the first part of this work, we discuss the theoretical basis of the new formula using common examples. In the second part, we introduce the more explicit form of this formula, thus allowing a more precise quantification of wettability by providing access to the direct measurement of liquid-solid adhesive forces. The third part proposes a method for measuring static surface tension without the adverse effects of the substrate.
\end{abstract}

\section*{General introduction}

Capillarity plays an important role in a wide range of scientific and industrial fields \cite{de1985wetting}, \cite{barozzi2017notes}, which explains the sustained research interest in this phenomenon. For more than two centuries, scientists have consistently identified limitations within classical theory \cite{adam1941physics}, \cite{brown1947fundamental}, \cite{shuttleworth1950surface}, \cite{chapin1959two}, \cite{berry1971molecular}. The inclusion of the concept of contact angle further complicates the problem \cite{erbil2014debate}. The underlying issue is increasingly recognized to be rooted in physical principles \cite{drelich2019contact}, rather than merely technical shortcomings \cite{miller1985interfacial}, \cite{marmur2006soft}. If the static aspects of the problem are not addressed properly, the dynamic behavior, as exemplified by the work of Washburn \cite{washburn1921dynamics}, will inevitably remain unresolved. This study focuses primarily on the investigation of capillary rise and fall. Although not aiming to provide an explicit review of the numerous studies that have explored this problem \cite{barozzi2017notes}, \cite{brown1947fundamental}, \cite{chapin1959two}, \cite{jurin1718ii}, \cite{rayleigh1899xxxvi}, \cite{bikerman1978capillarity}, \cite{mccaughan1987capillarity}, \cite{vorlicek1988capillarity}, \cite{scheie1989upward}, \cite{tabor1991gases}, \cite{mccaughan1992comment}, \cite{pellicer1995physical}, \cite{gennes2004capillarity}, \cite{roura2007contact}, \cite{sophocleous2010understanding}, \cite{rodriguez2010derivation}, \cite{liu2018jurin}, this work seeks to advance a more coherent understanding of the fundamental processes involved.
\section{First Section: surface tension}
\subsection{Introduction to the first section}

In the examination of the scientific literature pertaining to capillarity, one frequently encounters misunderstandings regarding the microscopic and macroscopic origins of surface tension. Two main perspectives, somewhat incompatible, emerge. One attributes the microscopic origin of surface tension to anisotropic forces arising from unstable liquid-gas bonds, as opposed to more stable liquid-liquid bonds. This view emphasizes a macroscopic mechanical force acting tangentially at the liquid's surface, creating tension in the liquid film \cite{brown1947fundamental} , \cite{berry1971molecular}, \cite{shuttleworth1950surface}, \cite{tabor1991gases}, \cite{defay1966surface}, \cite{marchand2011surface}, \cite{durand2021mechanical}. The second perspective avoids the mechanical approach and instead prefers the energetic interpretation of the phenomenon \cite{adam1941physics}, \cite{burdon2014surface}, \cite{bikerman2013surface}, \cite{woodruff1973solid}, \cite{shaw1970introduction}. By extending the study to phenomena involving the notion of contact angle, the situation becomes more confusing \cite{erbil2014debate}, \cite{drelich2019contact}. As Brown has rightly noted \cite{brown1971molecular}, perhaps everything has already been said. However, the challenge remains in identifying the theory that most effectively addresses these insights. We believe that this modest work may serve as a useful tool to draw the attention of scientists to the existence of a novel approach that could offer improved control over capillary phenomena.
\subsection{Analysis of the problem}
Consider a system where the three phases liquid,solid, and gas (vapor) (L/S/G) are in contact. It is important to note that in a system consisting solely of the liquid phase, the surface tension $(\gamma_{LL})$, arising from cohesive forces, is considered an internal force. This force acts to stabilize the system by minimizing its contact area. In contrast, the interfacial tensions $(\gamma_{LG})$ and $(\gamma_{LS})$, which result from the anisotropic adhesive forces between liquid and gas (vapor) and liquid and solid, respectively, are considered external forces. These forces increase the system’s energy by expanding the contact area of the liquid with the external environment. Gravity is also considered a destabilizing external force. Consider a liquid system that transitions from an initial equilibrium state to a final equilibrium state due to the influence of external constraints, such as capillary forces. The molecules of the liquid, due to their relative mobility, respond to the new equilibrium conditions by adopting the least energetically costly path. It is this intrinsic behavior of the molecules that allows the system to acquire the most stable energy state possible (the minimum possible contact area) when it adapts to the new conditions imposed by the external forces. Attributing a specific direction to the movement of liquid molecules may contradict the concept of their relative mobility and could also conflict with the aforementioned energy minimization principle. This raises important questions that require further exploration. We will address these questions by examining various examples in subsequent discussions.    
\subsubsection{Surface balance}
It is well known that in the absence of gravity, a liquid drop in static equilibrium assumes a  spherical shape, thereby maximizing the number of liquid-liquid (L/L) bonds and minimizing the number of liquid-gas (L/G) bonds. However, when gravity is present, the liquid drop requires external support to maintain equilibrium. The contact between the drop and the solid increases the number of liquid-solid (L/S) bonds and correspondingly decreases the number of solid-gas (S/G) bonds. Consider a liquid-solid-gas (L/S/G) capillary system at equilibrium state (1) (see Fig.1). Let the liquid mass consist of $N_{1}$ molecules in (L/L) contact, $n_{1} $ molecules in (L/S) contact, and $ m_{1} $ molecules in (L/G) contact. As the system transitions to equilibrium state (2) under external constraints, such as a support or a capillary, the configuration evolves to include $ N_{2} $ molecules in (L/L) contact, $ n_{2} $ molecules in (L/S) contact and $ m_{2} $ molecules in (L/G) contact. For a closed, non-reactive system, composed solely of liquid phase, the total number of molecules remains constant throughout the process, such that $ N_{2}+n_{2}+m_{2} $ = $N_{1} + n_{1} + m_{1}$. Defining $\Delta N = N_{2}- N_{1}$, $ \Delta m = m_{2}-m_{1} $ and $\Delta n = n_{2}- n_{1}$, we obtain,
\begin{equation}
\Delta N = -(\Delta n + \Delta m)
\end{equation}
This equation indicates that for a given number $\Delta N$ of (L/L) bonds dissociated/formed, an equivalent number $(\Delta n + \Delta m)$ of (L/S) and/or (L/G) bonds are formed/dissociated. Consider equal unit surface areas $U_{LL} = U_{LS} = U_{LG}$ and assuming equal surface densities ${\Delta U_{LL}}/{\Delta N} = {\Delta U_{LS}}/{\Delta n} = {\Delta U_{LG}}/{\Delta m}$, the surface balance can be expressed as follows:  
\begin{equation}
\Delta A_{LL} = -(\Delta A_{LS}+ \Delta A_{LG}) \:\:and\:\: \Delta A_{LS} = -\Delta A_{SG}
\end{equation}
Where $\Delta A_{LL}$, $\Delta A_{LS}$, and $\Delta A_{LG}$ represent the variations in the (L/L), (L/S), and (L/G) contact areas of the liquid. It can be stated that a variation in the (L/L) contact surface induces an equal variation in the (L/S) and/or (L/G) interfaces and vice versa. This balance holds regardless of the thickness of the layer considered, provided that the number of monomolecular layers is the same for the different surfaces and interfaces involved.
\begin{figure}[ht]
\centering
\includegraphics[width=0.72\linewidth]{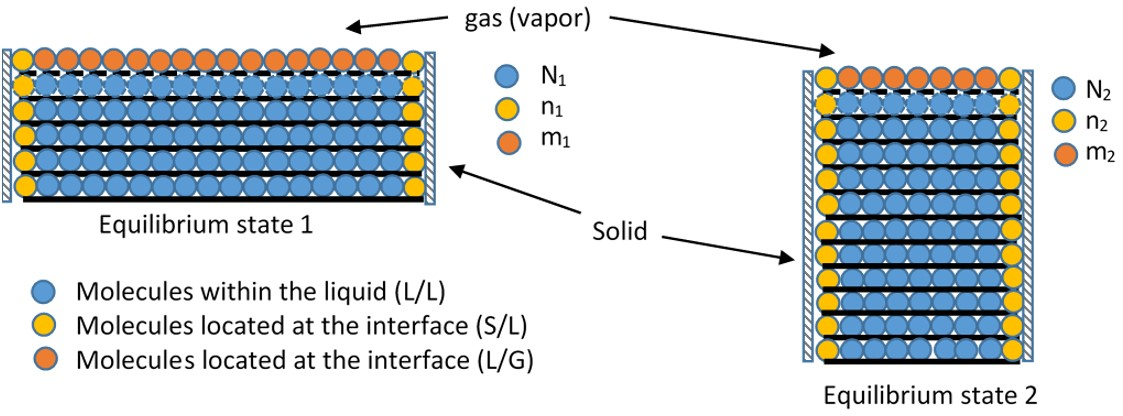}
\caption{\label{fig:Fig.1}Illustrates a liquid-solid-gas (L/S/G) system transitioning between two equilibrium states. }
\end{figure}
\subsubsection{Energy balance} The considered thickness is theoretically assumed to ensure homogeneous mean macroscopic values for the various surface forces, as derived from the expression for specific free energy $\gamma_{xy} = \left(\frac{\Delta F^{xy}}{\Delta A_{xy}}\right)_{T,V,n}$. Under the assumption of conservative forces, applying the principle of energy conservation to the system as it transitions between the two equilibrium states yields the following equation:
\begin{equation} (\gamma_{LS} - \gamma_{SG}) \Delta A_{LS} - \gamma_{LL} \Delta A_{LL} + \gamma_{LG} \Delta A_{LG} = W_{P} \end{equation}
$W_{P}$ represents the work of the gravitational forces. In Equation (A.3), the surface terms $A_{xy}$ are taken as positive. The sign preceding the specific energy term $(\gamma_{xy})$ indicates whether the conjugate surface is either formed (positive sign) or dissociated (negative sign). This energetic balance is valid for $\Delta N < 0$, where the work of adhesive forces (L/S) is dominant $(\gamma_{LS} > \gamma_{LL})$, as in the case of capillary rise, and for $\Delta N > 0$, where the work of cohesive forces is dominant $(\gamma_{LS} < \gamma_{LL})$, as in the case of capillary depression (glass/mercury system).

\subsection{Study of some common examples }
The most common examples used in the literature to discuss the origin of surface tension, as well as its direction, are illustrated in Fig. 2. For more details, see \cite{gennes2004capillarity} and \cite{adamson1967physical}.

\subsubsection{Study of Example (a) of Fig. 2}
In this example, we assume that the effect of the forces associated with the solid is negligible ($\Delta A_{LS} = 0$). Thus, Equation (A.2) simplifies to:
\begin{equation}
\Delta A_{LL} = -\Delta A_{LG}
\end{equation}
To maintain the liquid film in a stretched state ($\Delta A_{LG} > 0$), an external force $f$ is applied perpendicularly along the length $L$ of the rod. Given that $\left| \Delta A_{LL}\right| = \left| \Delta A_{LG}\right| = \Delta A$ and considering Equation (A.3), we can express the following:
\begin{equation}
(\frac{f}{L}+\gamma_{LG} -\gamma_{LL})\Delta A=W_{P}
\end{equation}
\begin{figure}[ht]
\centering
\includegraphics[width=0.72\linewidth]{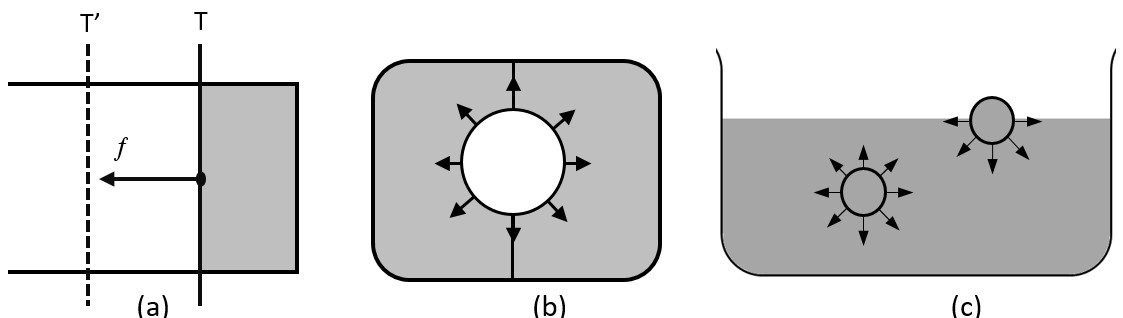}
\caption{\label{fig:Fig.2}Common examples used in the literature dealing with the origin of surface tension. }
\end{figure}
To interpret Equation (A.5), we refer to Fig. 3. In this figure, situations $(a_{i})$ depict the liquid film in its initial equilibrium state (position T), $(b_{i})$ show the film after spontaneous shrinkage (position T'), and $(c_{i})$ illustrate the film after forced stretching (position T'').
\begin{figure}[ht]
\centering
\includegraphics[width=1\linewidth]{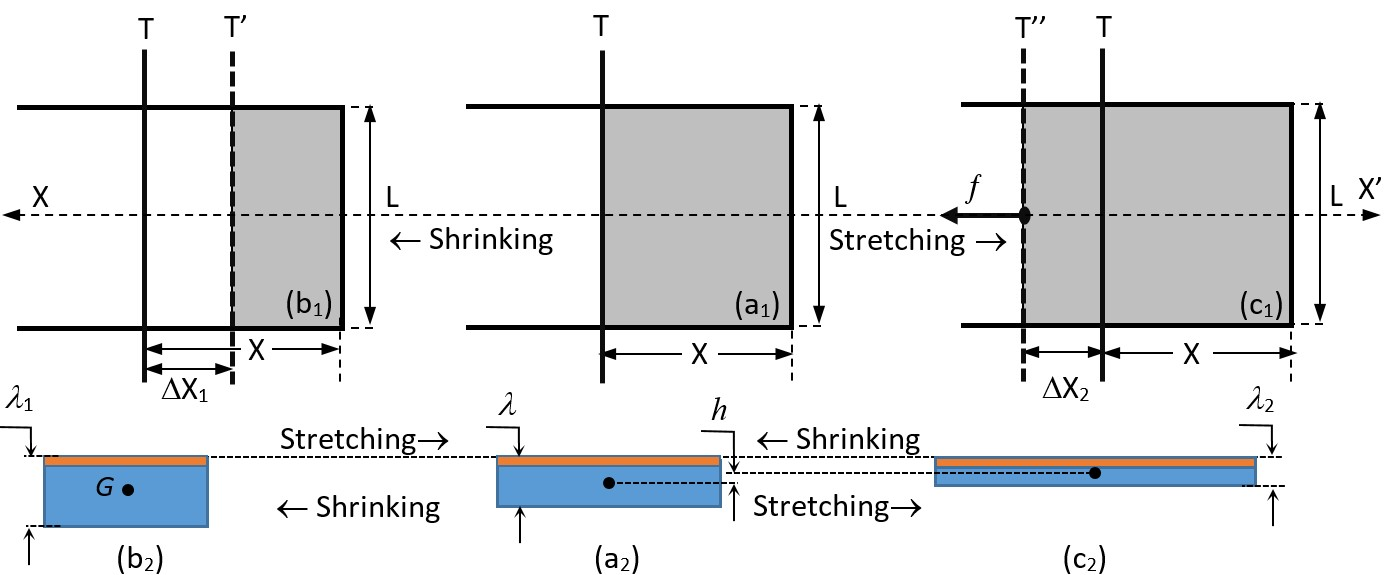}
\caption{\label{fig:Fig.3} Qualitative illustration of the phenomenon of stretching/shrinking of a liquid film.}
\end{figure}
Consider the scenario described in Cases $(3a_{i})$, where the liquid film is defined by its width $(L)$, length $(X)$, and uniform thickness $(\lambda)$. Under these conditions, the surface area of the film in contact with the gas is given by $A_{LG} = 2LX$, and its volume by $V = LX\lambda$. When the film undergoes an extension by a length $\Delta X_2$, the liquid-gas contact area increases by an amount $\Delta A_{LG} = 2L\Delta X_2$, as shown in cases $(3c_{i})$. For an incompressible volume, the relation $LX\lambda = L(X + \Delta X_2)\lambda_2$ holds, where the thickness changes from $\lambda$ to $\lambda_2$ as a result of stretching. This extension induces a displacement of the liquid’s center of mass by a distance $h = \left(\frac{\lambda - \lambda_2}{2}\right)$. Under these assumptions, Equation (A.5) can be rewritten as follows:
\begin{equation}
\frac{f}{L}=\gamma_{LL}-\gamma_{LG}+\Delta{\rho} g \frac{X}{\Delta X_2}\lambda\left(\frac{\lambda-\lambda_2}{4}\right)
\end{equation}
$\Delta{\rho}$ denotes the density difference between the liquid and vapor phases. To discuss the effect of weight, we simplify Equation (A.6) by assuming that the stretching has doubled the area of the film $(\Delta X_{2}=X)$. In this case, Equation (A.6) can be reduced to:
\begin{equation}
\frac{f}{L}=\gamma_{LL}-\gamma_{LG}+\Delta{\rho} g\lambda\left(\frac{\lambda-\lambda_2}{4}\right)
\end{equation}
This equation indicates that for very small thicknesses or minimal thickness variations $(\lambda_2\simeq \lambda)$, the effect of weight can be neglected, therefore:
\begin{equation}
\frac{f}{L}=\gamma_{LL}-\gamma_{LG}
\end{equation}
It is important to emphasize that Equation (A.8) incorporates two competing surface tensions, each arising from distinct physical interactions. The surface tension $\gamma_{LL}$ resulting from cohesive liquid-liquid forces, promotes contraction of the liquid-gas interface, while the surface tension $\gamma_{LG}$ associated with adhesive liquid-gas interactions, favors stretching of this interface. It can be clearly seen that Equation (A.8) provides a coherent alignment between mechanical and thermodynamic interpretations and reconciles microscopic and macroscopic viewpoints. This unifying aspect will be further clarified in the following sections.
\subsubsection{Energetic interpretation}
Cohesive energy, denoted as $(-E_{LL} > 0)$, is defined as the energy required to transfer liquid molecules from the liquid state (L/L) to the ideal gas state at a specified temperature and pressure \cite{berry1971molecular}. Similarly, adhesive energy, denoted as $(-E_{LG} > 0)$, corresponds to the energy required to move liquid molecules from the liquid-gas interface (L/G) to the ideal gas state. Stretching a liquid film involves transferring molecules from the (L/L) state to the (L/G) state. 
\begin{figure}[ht]
\centering
\includegraphics[width=0.8\linewidth]{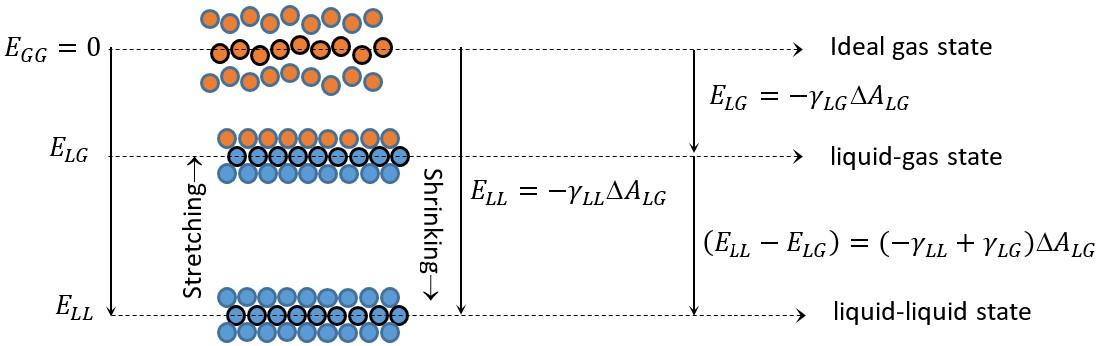}
\caption{\label{fig:Fig.4} Energy diagram of a stretched/shrunk film }
\end{figure}
As illustrated in Fig. 4, the energy needed to stretch the film by an amount $(\Delta A_{LG})$ is expressed as $[-E_{LL} + E_{LG} = (\gamma_{LL} - \gamma_{LG})\Delta A_{LG}]$. It is evident that a necessary condition for the existence of the liquid phase is $\gamma_{LL} > \gamma_{LG}$.
\subsection{Discussion of the direction of action of surface tension forces}
Discussing the direction of surface tension forces in a system in static or dynamic equilibrium, as in the case of example (c) in Fig. 2, is mechanically redundant, since a zero vector has neither magnitude nor direction. To determine the direction of a mechanical force acting on a system, it is essential to analyze the system’s behavior under that force, as illustrated by examples (a) and (b) in Fig. 2. For comparison with existing literature (see \cite{gennes2004capillarity}), we define $\gamma = \gamma_{LL} - \gamma_{LG}$, allowing Equation (A.8) to be rewritten as follows:
\begin{equation}
\frac{f}{L}=\gamma
\end{equation}
In this equation, the effects of liquid-solid interactions and gravity are not considered. Two key observations arise: first, $\gamma$ represents two opposing forces of different origins, suggesting that an energetic interpretation is more appropriate than a mechanical one. Second, the constraints imposed on the system in Example (a) of Fig. 2 prevent a generalized definition of the direction of $\gamma$. It is evident that the rod stretching the film has only one degree of freedom, along the $XX'$ direction, in which it can evolve, allowing the film to contract under the predominant effect of $\gamma_{LL}$. Thus, $\gamma_{LL}$ and consequently $\gamma$, act in the opposite direction to the force $f$, consistent with the principle of action and reaction.

According to the same equation, $\gamma_{LG}$ acts in the same direction as $f$. Thus, it is the constraints imposed on the system that determine the direction in which $\gamma$ responds. Imposing a predefined direction on $\gamma$ in all cases overlooks a key property of liquids—namely, the relative freedom of molecular mobility—and implicitly assumes that all scenarios are subject to similar constraints. Similarly, Example (b) in Fig. 2 is constrained to two degrees of freedom, minimizing the contact surface of the film. A broader definition must include a system under constraints that cause three-dimensional deformation. Capillary rise and fall exemplify this phenomenon, which will be explored in the following section.
\section{Second Section: capillary rise and fall phenomena}
\subsection{Introduction to the second section}
Unlike the case of thin homogeneous liquid films where surface deformations are considered two-dimensional, when a liquid rises or falls in a capillary, the surfaces and interfaces of the liquid undergo three-dimensional deformation. It is important to emphasize that the surface and energy balances derived in the first section apply to all capillary phenomena, including the rise and fall of liquids in capillaries. Consider the case of a narrow capillary with a radius $r$, in contact with a liquid’s surface contained in a wide container of radius $R$. Experimental observations show that, with glass and water, the liquid rises spontaneously in the capillary above the free surface of the liquid in the container. In contrast, with glass and mercury, the liquid descends. 
\subsection{Derivation of the capillary rise/fall formula}
Considering that only the water molecules within the volume enclosed by the dotted frame undergo changes in their energy state (as shown in the orange area in Fig. 5). The choice of the initial state as the reference state for establishing the energy balance allows for compensation of variations in the liquid-gas interfaces that occur when the liquid transitions from the container to the capillary (i.e., $\Delta A_{LG} = 0$). Under this assumption, the surface balance derived from Equation (A.2) simplifies to $\Delta A_{LS} = -\Delta A_{LL} = -\Delta A_{LG}$. By defining $|\Delta A_{LS}| = |\Delta A_{LL}| = |\Delta A_{SG}|$ as $\Delta A$ and applying Equation (A.3) to the entire system (capillary + container) as it evolves from the initial to the final state, as depicted in Fig. 5, we obtain the following equation:
\begin{equation}
(\gamma_{LS(r)}-\gamma_{SG}-\gamma_{LL})\Delta A^{cap}-(\gamma_{LS(R-r)}-\gamma_{SG}-\gamma_{LL})\Delta A^{cont}=W_{p}^{cap}
\end{equation}
\begin{figure}[ht]
\centering
\includegraphics[width=0.8\linewidth]{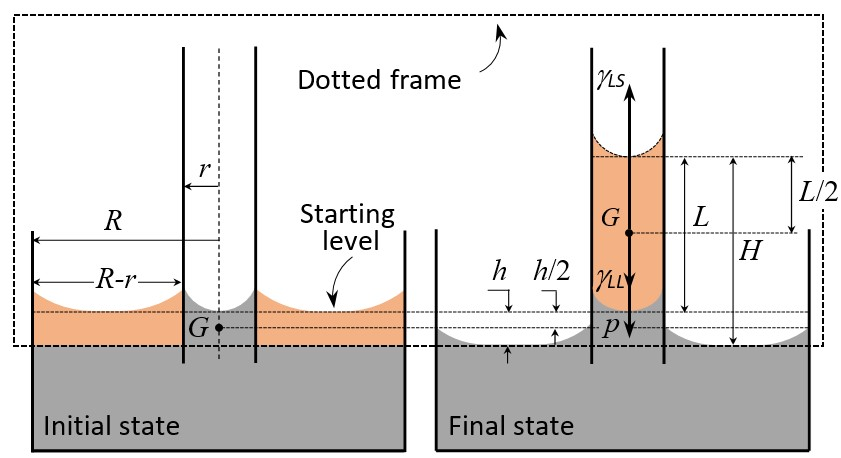}
\caption{\label{fig:Fig.5} Schematic of capillary rise. $R$ and $r$ represent the circular radii of the container and the capillary respectively. $H$ represents the height of the liquid column measured from the lowest point of the annular meniscus between the outer walls of the capillary and the inner walls of the container to the corresponding point in the capillary at the final equilibrium state. $(R-r)$ represents the thickness of the annular meniscus. $L$ represents the height of the liquid column measured from the initial starting level to the lowest point of the capillary meniscus in the final state. The parameter $h$ indicates the vertical displacement of the lowest point of the annular meniscus in the container between the initial and final states. $p$ denotes the weight of the liquid column of height $L$ within the capillary in the final state. Note that $H = L + h$.}
\end{figure}
Where $\Delta A^{cap}$ and $\Delta A^{cont}$ denote the increase in liquid contact area within the capillary and between the container's inner walls and the capillary's outer walls, respectively. As a first approximation, the contribution of the $\gamma_{SG}$ force can be neglected in comparison to the forces $\gamma_{LS}$ and $\gamma_{LL}$. Assuming that the heights of the annular meniscus on the outer wall of the capillary and on the inner wall of the container are identical, the following relationships can be written for the entire system:
\begin{equation}
(\gamma_{LS(r)}-\gamma_{LL})2\pi rL -(\gamma_{LS(R-r)}-\gamma_{LL})2\pi (R+r)h =\Delta\rho g\pi r^{2} L \frac{H}{2}
\end{equation}
The terms $(\gamma_{LS(r)} - \gamma_{LL})2\pi rL$ and $(\gamma_{LS(R-r)} - \gamma_{LL})2\pi(R + r)h$ correspond, respectively, to the contributions of the capillary forces acting within the capillary and the opposing capillary forces acting between the outer walls of the capillary and the inner walls of the container. The terms on the right-hand side of the equation represent the mechanical work required to raise the center of gravity $G$ of the liquid column in the capillary by a height $\frac{H}{2}$. Knowing that the liquid rising in the capillary is drawn from the container, then for an incompressible fluid, $[r^2 H = (R^2 - r^2) h]$, we
\begin{equation}
H=\frac{r^2}{(R^2-r^2)}L
\end{equation}
By substituting the value of $h$ from Equation (B.2) into Equation (B.3) and simplifying, the resulting equation becomes:
\begin{equation}
\gamma_{LS(r)}-\gamma_{LL}-(\gamma_{LS(R-r)}-\gamma_{LL})\frac{r}{R-r} =\frac{\Delta\rho gr}{2}\frac{H}{2}
\end{equation}
We see from Equation (B.4) that the term representing the capillary forces relative to the container,
$(\gamma_{LS(R-r)}-\gamma_{LL})\left(\frac{r}{R-r}\right)$  becomes negligible as the capillary narrows $(r\rightarrow 0)$ and/or as the container widens $(R\rightarrow \infty)$.
\subsubsection{Discussion of the Case When Adhesive Forces Dominate Cohesive forces (e.g., the Glass/Water System)}
When adhesive forces dominate cohesive forces $(\gamma_{LS((R-r),r)}>\gamma_{LL})$, three main cases arise.\\ \textbf{Case (6.a)} pertains to the situation where the capillary is sufficiently narrow and the container sufficiently wide $R \gg r$. In this case, the term $\left(\frac{r}{R-r}\right)$ approaches zero, allowing us to neglect the term $(\gamma_{LS(R-r)}-\gamma_{LL})\left(\frac{r}{R-r}\right)$. Consequently, Equation (B.4) can be simplified to the following form:
\begin{equation}
H=\frac{4(\gamma_{LS(r)}-\gamma_{LL})}{\Delta\rho gr}
\end{equation}
We can see from Equation (B.5) that, for $\gamma_{LS(r)} > \gamma_{LL}$, $H$ is a positive quantity, indicating that the liquid rises in the capillary. This result arises from the competition between adhesive forces on the one hand and the combined effects of cohesive forces and gravitational forces on the other. The force diagram based on Equation (B.5) is plotted in Fig. 5. An important question may arise: at what value of $R$ does the influence of the container on the liquid column in the capillary become exactly zero? The answer to this question can be determined by analyzing the capillary forces acting between the outer walls of the capillary and the inner walls of the container, namely, the term $\left(\gamma_{LS(R-r)} - \gamma_{LL}\right)\frac{r}{R-r}$ in Equation (B.4). This point will be discussed in detail in Part 2 of this work, where the explicit forms of the forces $\gamma_{LS(R-r)}$ will be introduced. Furthermore, in this context, the influence of the submerged portion of the capillary on the height of the liquid rise, as explored in certain studies \cite{extrand2016origins}, \cite{maity2019capillary}, will also be examined.
\begin{figure}[ht]
\centering
\includegraphics[width=0.8\linewidth]{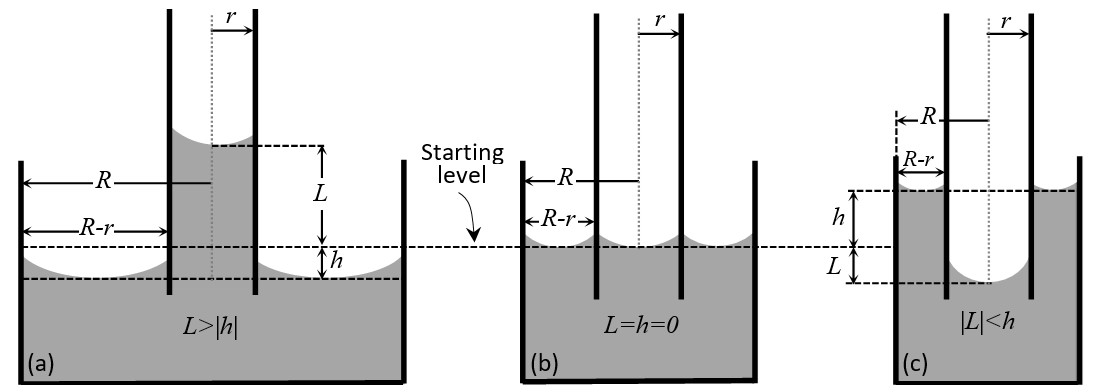}
\caption{\label{fig:Fig.6} Illustration of the three primary cases of capillary rise, arising from the competition between the adhesive forces of the capillary and those of the container. }
\end{figure} \\
\textbf{Case (6.b)} corresponds to the scenario where the level of liquid within the capillary is equal to that in the surrounding container. This situation implies that, after the capillary is introduced, no net energy exchange occurs between the capillary and the container. Under these conditions, the capillary forces inside the capillary are exactly balanced by those exerted between the outer walls of the capillary and the inner walls of the container. Theoretically, this equilibrium is reached when $R = 2r$. In this case we have $(H=L = h = 0)$, and Equation (B.4) simplifies as follows:
\begin{equation}
\gamma_{LS(R-r)}=\gamma_{LS(r)}
\end{equation}
\textbf{Note}: In practice, the expressions for $\gamma_{LS((R-r),r)}$ depend, among other factors, on the shape of the cross-section of the container that holds the liquid. For more information, refer to Part 2.
\\ \textbf{Case (6.c)} refers to the scenarios in which the capillary radius approaches that of the container
$(r\rightarrow R)$. As a result, the ratio $\frac{R-r}{r}$ tends to zero, allowing the capillary forces exerted inside the capillary $(\gamma_{LS(r)}-\gamma_{LL})\frac{R-r}{r}$ to be neglected in favor of the forces exerted between the outer capillary walls and the inner container walls $(\gamma_{LS(R-r)}-\gamma_{LL})$. Consequently, Equation (B.4) simplifies to:
\begin{equation}
H=\frac{-4(\gamma_{LS(R-r)}-\gamma_{LL})}{\Delta\rho g(R-r)}
\end{equation}
Given that \(\gamma_{LL(R-r)} > \gamma_{LL}\) and that $R > r$, the right-hand side of Equation (B.7) is negative, implying that $H$ is a negative quantity. This indicates a decrease in the liquid level in the capillary. 

\subsubsection{Discussion of the case when the cohesive forces dominate the adhesive forces (e.g. the glass/mercury system)}
In cases where cohesive forces dominate over adhesive forces, Equation (B.4) remains valid. Fig. 7 illustrates the three main limiting cases.
\begin{figure}[ht]
\centering
\includegraphics[width=0.8\linewidth]{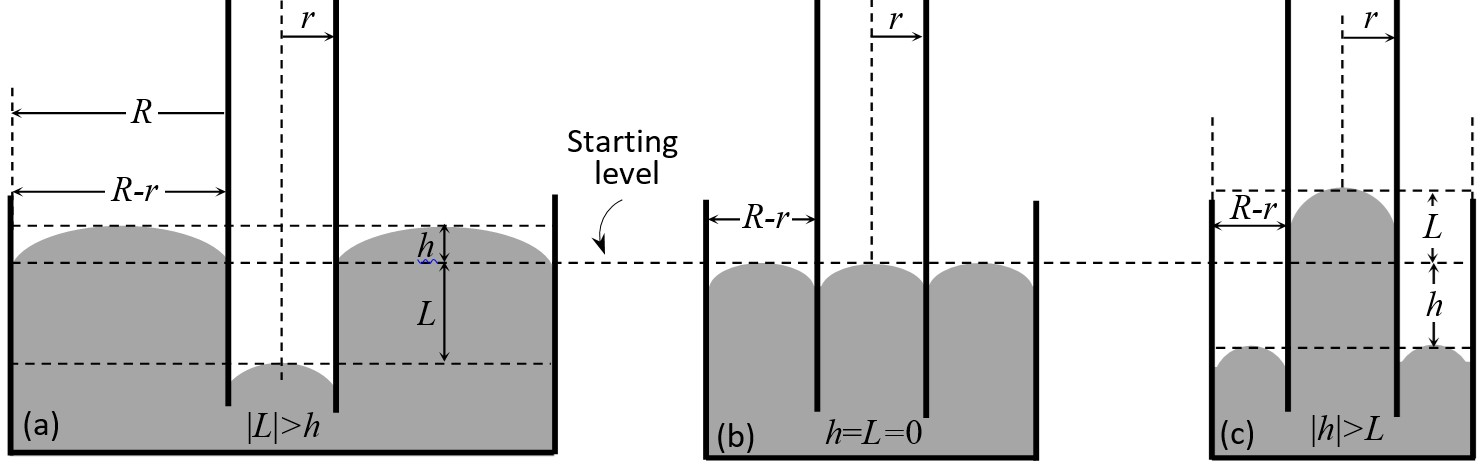}
\caption{\label{fig:Fig.7} Illustration of the three primary cases of capillary depression. The radius of the annular meniscus (R-r) decreases from left to right.}
\end{figure} \\
\textbf{Case (7.a)} is discussed analogously to Case (6.a), where the container is sufficiently large. Equation (B.5) shows that for $\gamma_{LS(r)} < \gamma_{LL}$, $H$ is a negative value, indicating that the liquid descends within the capillary. The force diagram corresponding to this situation is presented in Fig. 8.\\
\begin{figure}[ht]
\centering
\includegraphics[width=0.78\linewidth]{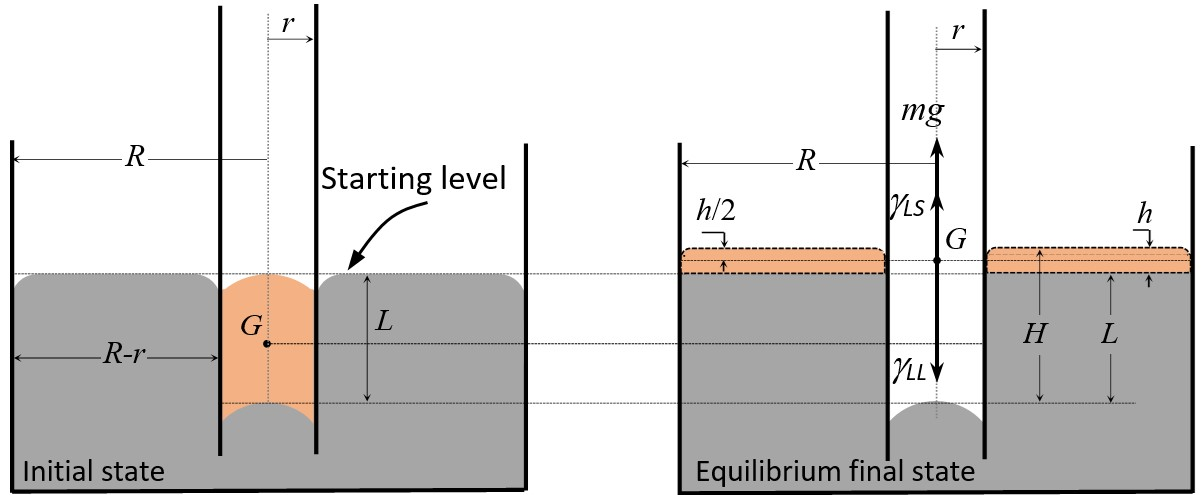}
\caption{\label{fig:Fig.8} Detailed scheme of scenario (7.a) where $(R\gg2r)$} 
\end{figure}
\textbf{Case (7.b)} is similar to case (6.b), as it describes a situation where the capillary forces are exactly counterbalanced by the container forces, resulting in no level difference between the two surfaces $(L = h = 0)$. This scenario is theoretically achieved when $(R = 2r)$ and aligns with the results of Equation (B.6).
\\ \textbf{Case (7.c)} is analogous to Case (6.c), as both describe scenarios where the container is sufficiently wide, and Equation (B.7) remains valid. The primary distinction between the two cases is that in Case (7.c), we have $\gamma_{LS(R-r)} < \gamma_{LL}$. Consequently, for $(R-r) > 0$, the right-hand side of Equation (B.7) is positive, which implies that $H$ is also positive, indicating that the liquid rises within the capillary. This phenomenon can be readily observed experimentally. Furthermore, it should be noted that, in practice, the absolute value of the right-hand side of the equation increases with $r$, further supporting the results (see Part 2 of this work).

\section*{Concluding remarks}
In this first part, we have detailed the physical framework underlying our equations by analyzing the most fundamental cases. It is clearly evident that the proposed formula is based on reasoning that follows a logical and coherent structure, maintaining consistency across scales, from the microscopic to the macroscopic level, and throughout all the examples discussed. The proposed formula, as a single, fundamental expression without arbitrary parameters, demonstrates an unprecedented capacity to qualitatively describe a wide range of scenarios, including the behavior of liquid films and the main limiting cases of capillary rise and fall. These scenarios include cases where adhesive forces dominate cohesive forces, as in the glass/water system, and where cohesive forces dominate adhesive forces, as in the glass/mercury system.
The quantitative analysis of this problem will be the focus of Part 2 of this study, where the explicit form of the $\gamma_{LS((R-r),r)}$ forces will be introduced. The introduction of the concept of the "capillary range" extends the applicability of this formula to various cross-sectional capillary geometries (rectangular, square, triangular, etc.) and facilitates the resolution of key challenges, such as quantifying wettability. This is achieved by enabling the direct experimental measurement of short-range adhesive forces $\Gamma_{LS}$, a critical parameter of interest.
This new theoretical framework explicitly addresses fundamental questions, such as: How does $\gamma_{LS(R-r)}$ vary and what is its relationship to the shape of the meniscus (contact angle)? At what critical radius of the container, $(R)$, does the influence of the container wall on the height of the liquid column in the capillary vanish? At what critical radius of the capillary, $(r)$, does the liquid begin to rise within the capillary?
Furthermore, this formula enables the calculation of the liquid volume that rises along the sides and corners of containers under both standard gravity and microgravity conditions, for various cross-sectional shapes and for both open and closed container bases. It enhances the understanding of capillary pressure and allows for more precise quantification. The expression for the Bond number is refined and the hysteresis phenomenon is effectively explained and quantified.  

\section{Acknowledgments}
The author thanks Dr. Deghmoum Abdelhakim for his linguistic assistance.
\printbibliography
\end{document}